\documentstyle[prc,aps,graphicx,preprint,tighten]{revtex}

\newcommand{\siml}{\raisebox{-.5ex}{$\stackrel{>}{\sim}$}}

\begin{document}
\title{
Is early thermalization achieved only near midrapidity at RHIC ?
}
\draft
\author{Tetsufumi Hirano\footnote{E-mail: hirano@nt.phys.s.u-tokyo.ac.jp}}
\address{Physics Department, University of Tokyo, Tokyo 113-0033, Japan}
\date{\today}
\maketitle
\begin{abstract}
The pseudorapidity dependence of elliptic flow in Au+Au collisions at 130 $A$ GeV is studied within a full three-dimensional hydrodynamic model in the light-cone coordinate.
First, we prepare two initial conditions in the hydrodynamic model for analyzing elliptic flow.
Both initial conditions lead to reasonable agreement with single particle spectra in central and semi-central collisions.
Next, by using these hydrodynamic simulations, we compare elliptic flow as a function of pseudorapidity with experimental data recently measured by the PHOBOS Collaboration.
Our results are in agreement with experimental data only near midrapidity.
This suggests that thermalization in the early stage of collisions is not achieved in forward and backward rapidity regions.
\end{abstract}

Nucleus-nucleus collisions at the Relativistic Heavy Ion Collider (RHIC) give us an opportunity to study a new state of deconfined nuclear matter, the quark-gluon plasma (QGP) \cite{QM2001}.
The main goals in the physics of relativistic heavy-ion collisions are not only the discovery of the QGP, but also the investigation of thermodynamical aspects of its new phase, i.e., the equation of state (EOS), the order of phase transition between the QGP phase and the hadron phase, or the critical temperature. 
It is very important to check whether the thermalization of the nuclear matter produced at the very early stage of collisions is really achieved, before discussing the thermodynamics of the QGP.
Elliptic flow in non-central collisions \cite{OLLITRAULT} is suited for this purpose.
If produced particles frequently rescatter with each other, we naively expect that thermalization of the system is achieved and that the large pressure is built in the reaction zone.
The pressure produces the momentum anisotropy of observed particles from the spatial deformation in the transverse plane.
The coefficients of second harmonics in the azimuthal distribution, not only its magnitude $v_2$ but also its transverse momentum dependence $v_2(p_t)$ and its centrality dependence $v_2(b)$, seem to be good indicators for thermalization of nuclear matter \cite{SORGE,HEISEL,ZHANG,VOLO00,KOLB,TEANEY,ZABRODIN}.
Hydrodynamic simulations \cite{KOLB}, in which full local thermalization is assumed, give us excellent agreement with the first result of $v_2(p_t)$ from Au+Au 130 $A$ GeV collisions at the RHIC \cite{STAR} up to $p_t\sim1.5$ GeV near midrapidity.
Their results are based on a (2+1)-dimensional hydrodynamic model with the Bjorken's scaling solution.
One cannot, however, discuss the rapidity dependence of observables by exploiting the Bjorken's model \cite{BJOR}.
Since this model is assumed to give a good description of space-time evolution of nuclear matter near the midrapidity region, the agreement between the model calculation and experimental data means that early thermalization at the RHIC is achieved \textit{at least only at midrapidity}.
In this paper, we investigate how far from midrapidity thermalization in the early stage is achieved through comparison the pseudorapidity dependence of elliptic flow from a genuine three-dimensional hydrodynamic model with experimental data recently measured by the PHOBOS Collaboration \cite{PARK}.

There are several works based on \textit{full} three-dimensional hydrodynamic simulations for one fluid \cite{AMELIN,RHLLE,NONAKA,HIRANO,OSADA}.
These simulations, except Ref. \cite{OSADA}, are performed in the Cartesian coordinate.
In view of numerical analysis, hydrodynamic simulations in the Cartesian coordinate at the RHIC energy are tremendously hard owing to very long life time of the fluid ($t_f\sim100$ fm).
So we extend our previous algorithm in the Cartesian coordinate \cite{HIRANO} to the `light-cone' (or Bjorken) coordinate for saving of the computation time.
In the new coordinate, hydrodynamic equations ($\partial_\mu T^{\mu\nu} = 0$ and $\partial_\mu n_{\mathrm{B}}^\mu=0$) are written as
\begin{eqnarray}
\label{HE}
\partial_\tau \left(\begin{array}{c}
             U_1 \\
             U_2 \\
             U_3 \\
             U_4 \\
             U_5 \\ 
            \end{array} \right)
& + & \nabla \cdot \left(\begin{array}{c}
             U_1 \\
             U_2 \\
             U_3 \\
             U_4 \\
             U_5 \\ 
            \end{array} \right) \tilde{{\bf v}}
+  \left(\begin{array}{c}
             \tau \partial_x P \\
             \tau \partial_y P \\
             \partial_{\eta_{\mathrm{s}}} P \\
             \tau \nabla \cdot P \tilde{{\bf v}}\\
             0\\ 
            \end{array} \right)
+  \left(\begin{array}{c}
             0 \\
             0 \\
             U_3/\tau \\
             U_4 \tilde{v}_{\eta_{\mathrm{s}}}^2/\tau+P(1+\tilde{v}_{\eta_{\mathrm{s}}}^2)\\
             0\\ 
            \end{array} \right) = 0,
\end{eqnarray}
where,
\begin{eqnarray}
\label{U}
\left(\begin{array}{c}
             U_1 \\
             U_2 \\
             U_3 \\
             U_4 \\
             U_5 \\ 
            \end{array} \right) & = &
\left(\begin{array}{c}
             \tau \tilde{\gamma}^2 (E+P)\tilde{v}_x \\
             \tau \tilde{\gamma}^2 (E+P)\tilde{v}_y \\
             \tau \tilde{\gamma}^2 (E+P)\tilde{v}_{\eta_{\mathrm{s}}} \\
             \tau \tilde{\gamma}^2 (E+P)-\tau P \\
             \tau \tilde{\gamma} n_{\mathrm{B}} \\ 
            \end{array} \right).
\end{eqnarray}
Here $E$, $P$, and $n_{\mathrm{B}}$ are energy density, pressure, and baryon density;
$\tau = \sqrt{2x_+ x_-}=\sqrt{t^2-z^2}$, $\eta_{\mathrm{s}}=(1/2)\log(x_+/x_-)=(1/2)\log[(t+z)/(t-z)]$, $x$, and $y$ are, respectively, the proper time, the space-time rapidity, the transverse coordinate parallel to the impact parameter vector, and the coordinate perpendicular to the reaction plane.
Note that $\nabla = (\partial_x, \partial_y, \partial_{\eta_{\mathrm{s}}}/\tau)$.
Fluid velocities in the new coordinate are represented in terms of $v_x$, $v_y$, and $v_z$;
$\tilde{v}_x  =  v_x \cosh Y_{\mathrm{f}}/\cosh(Y_{\mathrm{f}}-\eta_{\mathrm{s}})$, 
$\tilde{v}_y  =  v_y \cosh Y_{\mathrm{f}}/\cosh(Y_{\mathrm{f}}-\eta_{\mathrm{s}})$, and $\tilde{v}_{\eta_{\mathrm{s}}}  =  \tanh(Y_{\mathrm{f}}-\eta_{\mathrm{s}})$, where $Y_{\mathrm{f}}  =  (1/2)\log[(1+v_z)/(1-v_z)]$ is the rapidity of a fluid element. The Lorentz gamma factor is $\tilde{\gamma} = 1/\sqrt{1-\tilde{{\bf v}}^2}$.
We use the same model EOS as the one represented in Ref.\cite{NONAKA}.
The EOS has a first order phase transition between the QGP phase and the hadron phase at $T_{\mathrm{c}}(n_{\mathrm{B}}=0)=160$ MeV.
The QGP phase is assumed to be free gas composed of quarks with $N_f = 3$ and gluons. For the hadron phase we adopt a resonance gas model, which includes all baryons and mesons up to the mass of 2 GeV \cite{PDG}, together with an exclude volume correction \cite{XV}.
For further details on the EOS, see Refs.\cite{NONAKA,HIRANO}.

We set up initial conditions at $\tau_0 = 0.6$ fm.
The initial energy density is motivated by a tilted disc recently discussed in the context of `anti-flow' or `third flow component' \cite{CSERNAI,BRACHMANN}. 
We assume that the longitudinal profile of initial energy density $E(\eta_s)$ at a transverse coordinate ($x$, $y$) is composed of two regions; the initial energy density is flat near $\eta_s \sim 0$ and smoothly connects to vacuum as a half part of a Gaussian function in the forward and backward space-time rapidity regions. The length of a flat region $\Delta \eta_{\mathrm{flat}}$ and the width of Gaussian function $\Delta \eta_{\mathrm{Gauss}}$ are adjustable parameters to be determined by the experimental data of the (pseudo)rapidity distribution.
In symmetric collisions with zero impact parameter, we expect $E(\eta_s)=E(-\eta_s)$.
On the other hand, we shift the energy density by $\Delta \eta_{\mathrm{s}}$ which is identified with the center-of-rapidity for each transverse coordinate \cite{SOLL} in non-central collisions.
The difference of thickness between two colliding nuclei for each transverse coordinate results in the initial distribution like a tilted disc.
We compare two initial conditions in this paper.
In order to reproduce a dip structure of the pseudorapidity distribution observed by the PHOBOS Collaboration\cite{PHOBOS}, we choose $\Delta \eta_{\mathrm{flat}}=6.0$, $\Delta \eta_{\mathrm{Gauss}}=0.4$, and a maximum energy density $E_{\mathrm{max}} = 40$ GeV/fm$^3$ for the initial condition \textit{A} (IC \textit{A}).
The other initial condition (IC \textit{B}) has a smaller flat region $\Delta \eta_{\mathrm{flat}}=2.8$, a larger Gaussian width $\Delta \eta_{\mathrm{Gauss}}=1.6$, and a slightly larger maximum energy density $E_{\mathrm{max}} = 43$ GeV/fm$^3$ than the IC \textit{A}. The latter also gives a reasonable result of the pseudorapidity distribution.
Both initial conditions for energy density at $\eta_{\mathrm{s}}=0$ are the same as the model `eBC' in Ref.\cite{KOLB2}, i.e., the transverse profile of energy density is in proportion to the number of binary collisions in the transverse plane.
This model gives the similar shape of the charged particle yield per participating nucleon pair $(dN_{\mathrm{ch}}/d\eta)/(0.5N_{\mathrm{part}})$ as a function of $N_{\mathrm{part}}$ \cite{KOLB2}.
Figure 1 shows two sets of initial energy density in the reaction plane.
We adopt the following initial baryon density and flow velocities as common initial conditions.
For the initial baryon density, we extend the parameterization by Sollfrank \textit{et al.} \cite{SOLL}, originally proposed for central collisions, to the case for non-central collisions.
We found that the shape of baryon density does not largely affect the final results obtained in this paper since the produced matter at the RHIC energy is dominated by mesons.
Initial longitudinal flow is the Bjorken's solution $\tilde{v}_{\eta_{\mathrm{s}}}(\tau_0)  =  0$ (or $Y_{\mathrm{f}}(\tau_0) = \eta_{\mathrm{s}}$) and transverse flow velocities vanish $\tilde{v}_x(\tau_0) = \tilde{v}_y(\tau_0) = 0$.

Figure 2 shows the pseudorapidity distribution of charged particles for central and semi-central collisions.
We accumulate the contribution from particles directly emitted from freeze-out hypersurface and feeding from resonance decays.
The preliminary experimental data are given by the PHOBOS Collaboration \cite{PHOBOS}.
The PHOBOS Collaboration estimates the average numbers of participants $<N_{\mathrm{part}}> \sim 340$ for the 0-6\% central collisions and $\sim 93$ for the 35-45\% semi-central collisions \cite{PHOBOS}.
We choose the impact parameters 2.4 fm for central and 8.9 fm for semi-central collisions.
The resultant numbers of participants based on the wounded nucleon model with the standard Woods-Saxon profile are 342 ($b=2.4$ fm) and 94 ($b=8.9$ fm), respectively.
We first tune initial parameters to reproduce central events.
By changing the impact parameter and unchanging the other parameters, we also obtain the results for another centrality.
Although the IC \textit{A} is the best fit to experimental data, the IC \textit{B} is also acceptable within error bars.
It should be noted that a dip structure at $\eta=0$ results from the Jacobian of the transformation from rapidity $Y$ to pseudorapidity $\eta$ \cite{KOLB3}.

In order to describe the dominant radial flow at freeze-out, we choose an appropriate value for freeze-out energy density so as to reproduce the slope of transverse momentum distributions.
In Fig. 3 we compare $p_t$ spectra of negative pions and anti-protons for central and semi-central collisions with experimental data measured by the PHENIX Collaboration \cite{PHENIX}.
By choosing the freeze-out energy density $E_{\mathrm{f}} = 120$ MeV/fm$^3$ which corresponds to the mean freeze-out temperature $\left.<T_{\mathrm{f}}>\right|_{\eta_{\mathrm{s}}=0} \sim 137$ MeV, our results are in very good agreement with the slope of these spectra.
The resultant mean radial flow $\left.<v_{\mathrm{r}}>\right|_{\eta_{\mathrm{s}}=0}$ are 0.49 ($b=2.4$ fm) and 0.41 ($b=8.9$ fm) for the IC \textit{A} and 0.50 ($b=2.4$ fm) and 0.42 ($b=8.9$ fm) for the IC \textit{B}.

Finally we perform hydrodynamic simulations with various impact parameters  and obtain the pseudorapidity dependence of elliptic flow $v_2(\eta)$ for charged particles in minimum bias events
\begin{eqnarray}
v_2(\eta)=\frac{\displaystyle \int bdb d\phi \cos2\phi\frac{d^2 N_{\mathrm{ch}}}{d\eta d\phi}(b)}{\displaystyle \int bdb\frac{dN_{\mathrm{ch}}}{d\eta}(b)}.
\end{eqnarray}
Our results and experimental data obtained by the PHOBOS Collaboration are shown in Fig. \ref{FIG4}.
For the IC \textit{A}, two bumps appear at $| \eta | \sim 3$.
This structure is caused by the tilted disc shape, i.e., produced matter is highly deformed in forward and backward space-time rapidity regions  (see also Fig. 1 (a)).
The initial energy density for the IC \textit{A} has a `crescent' shape rather than an almond shape at $| \eta_{\mathrm{s}} | \sim 3.5$ in the transverse plane \cite{BRACHMANN}.
Since the initial deformation of energy density for the IC \textit{B} is almost independent of $\eta_{\mathrm{s}}$ owing to a relatively large $\Delta \eta_{\mathrm{Gauss}}$ (see also Fig. 1 (b)), the resultant $v_2(\eta)$ has no bumps.
Thus we find the pseudorapidity dependence of elliptic flow is highly sensitive to the initial longitudinal profile of energy density.
In both cases, our results are in reasonable agreement with experimental data only near the midrapidity region.
These results indicate that thermalization is fully achieved near the midrapidity region and partially in forward and backward regions.
We note that our $v_2(p_t)$ and $v_2(n_{\mathrm{ch}}/n_{\mathrm{max}})$ in $| \eta | < 1.3$ (not shown) are consistent with the results shown in Ref.\cite{KOLB}; our results are consistent with the STAR data \cite{STAR} below $p_t \sim 1$ GeV and above $n_{\mathrm{ch}}/n_{\mathrm{max}} \sim 0.5$.

Three discussions are in order here:

(i) It is best to compare elliptic flow from hydrodynamics with experimental data for each centrality in order to see how the thermalized region increases with centrality.
In addition to elliptic flow, we need to analyze the transverse momentum (mass) distribution in the forward rapidity region since radial flow is also appropriate for checking thermalization of the system.
There are no such data at present.
Therefore we hope the forthcoming experimental data of $v_2(\eta)$ for each centrality or the transverse momentum distribution in the forward rapidity region will reveal the dynamics of nuclear matter in the whole phase space.

(ii) It is assumed in conventional hydrodynamic models that the sharp transition from the hydrodynamic piture to the non-interacting particle picture happens at the constant temperature.
On the other hand, the sophisticated freeze-out procedure yields small values of $v_2$ even if thermalization sets in the early stage of collisions \cite{TEANEY}.
The model evolves the QGP and mixed phase as a relativistic fluid, while it switches to a hadronic cascade model (RQMD) at a temperature $T_{\mathrm{switch}}=160$ MeV.
The main advantage of this hybrid model is that the freeze-out process is automatically included without any assumption such as freeze-out temperature.
This hybrid model can give small $v_2$ in forward or backward rapidity regions where the multiplicity or the number of hadronic rescattering is small.

(iii) There may exist other initial conditions for hydrodynamic simulations which lead to be in agreement with single particle spectra and elliptic flow.
Along the lines of thought, more sophisticated initialization based on the nuclear thickness function will be studied elsewhere \cite{PASI}. 

In summary, we have studied early themalization of produced particles at the RHIC by using a genuine three-dimensional hydrodynamic model.
Our hydrodynamic simulations are performed in the light-cone coordinate.
Such simulations become important in analyzing ultra-relativistic heavy-ion collisions at the RHIC energies.
We considered two sets of initial parameters in the hydrodynamic model.
We chose initial parameters so as to reproduce the pseudorapidity and transverse momentum spectra for both central and semi-central collisions.
By using results from these numerical simulations, we analyzed the pseudorapidity dependence of elliptic flow.
We found that a large value of $v_2$ comes from the tilted disc shape which has  a large spatial deformation in the transverse plane in the forward rapidity region.
Near $\eta = 0$ our results for minimum bias events are consistent with experimental data observed by the PHOBOS Collaboration.
On the other hand, we failed to reproduce experimental data in forward and backward rapidity regions ($| \eta | \siml 1$).
This indicates the early pressure is not built in forward and backward regions.
Both theoretical and experimental detailed analyses as discussed above are indispensable for understanding the dynamics of nuclear matter not only at midrapidity but also at forward (backward) rapidity at the RHIC energy.

The author is much indebted to S. Esumi, T. Hatsuda, P. Huovinen, Y. Miake, K. Morita, S. Muroya, and C. Nonaka for fruitful discussions.
He also thanks B. Back, S. Manly, and I. Park for teaching experimental aspects to him and C. Nonaka for providing him with a numerical table of EOS.
He also acknowledges Inoue Foundation for Science for the financial support.


%
%
 \begin{figure}
 \begin{center}
 \includegraphics[width=0.8\textwidth]{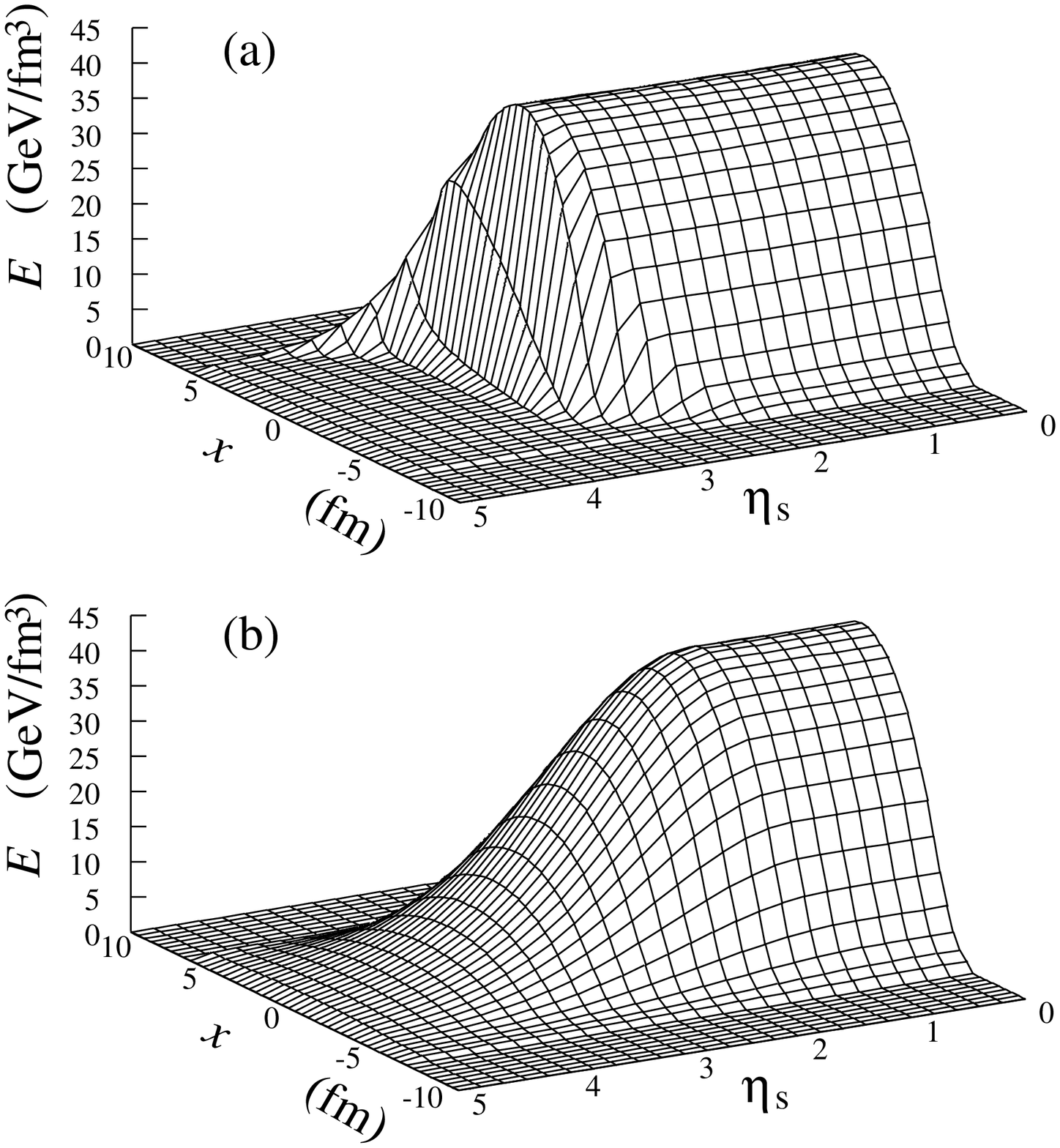}
 \end{center}
 \caption{Two sets of initial energy density in the reaction plane ($y=0$ and $\eta_{\mathrm{s}}>0$). (a) Initial condition \textit{A}. (b) Initial condition \textit{B}. See text for details.}
 \label{FIG1}
 \end{figure}

 \begin{figure}
 \begin{center}
 \includegraphics[width=0.8\textwidth]{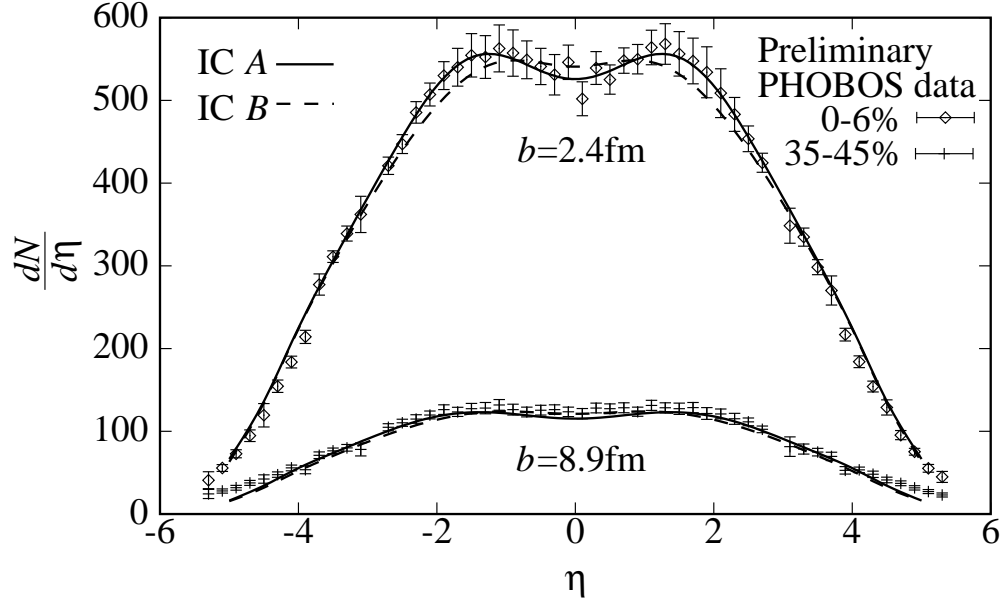}
 \end{center}
 \caption{Pseudorapidity distribution of charged particles in Au+Au 130 $A$ GeV central and semi-central collisions. Solid lines and dashed lines correspond to initial conditions \textit{A} and \textit{B}, respectively. Experimental data are observed by the PHOBOS Collaboration.}
 \label{FIG2}
 \end{figure}

 \begin{figure}
 \begin{center}
 \includegraphics[width=0.8\textwidth]{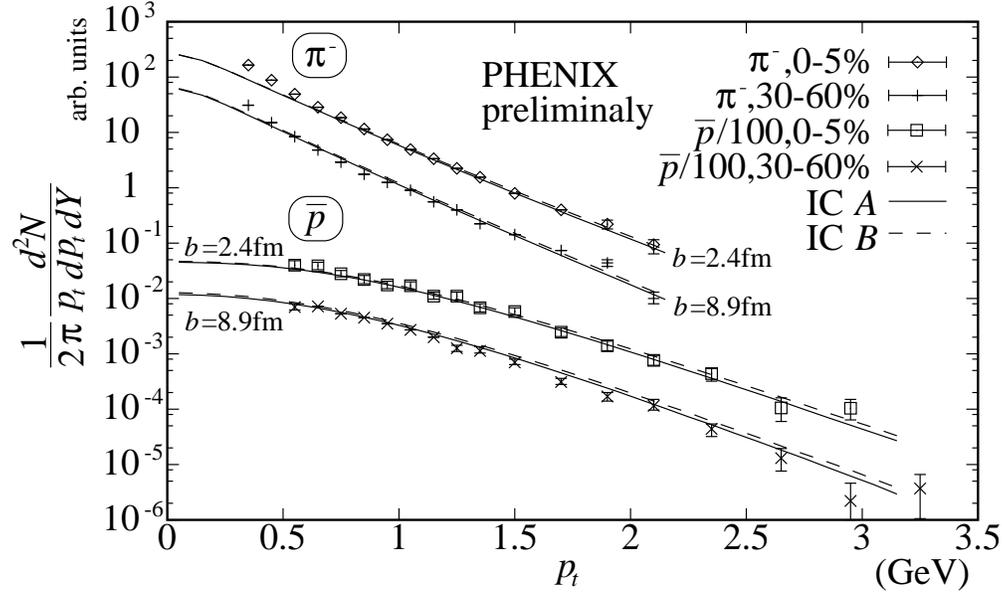}
 \end{center}
 \caption{Scaled transverse momentum distribution of negative pions and anti-protons in Au+Au 130 $A$ GeV central and semi-central collisions. Solid lines and dashed lines correspond to initial conditions \textit{A} and \textit{B}, respectively. Experimental data are observed by the PHENIX Collaboration.}
 \label{FIG3}
 \end{figure}

 \begin{figure}
 \begin{center}
 \includegraphics[width=0.8\textwidth]{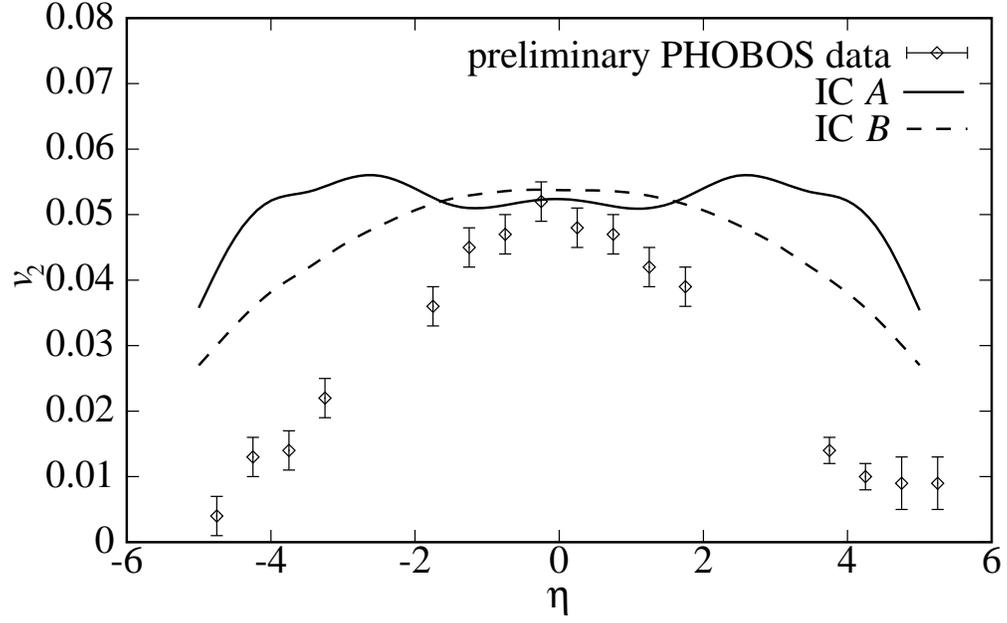}
 \end{center}
 \caption{Pseudorapidity dependence of elliptic flow for charged particles in Au+Au 130 $A$ GeV collisions. The value of elliptic flow is averaged over all centrality. Solid line and dashed line correspond to initial conditions \textit{A} and \textit{B}, respectively. Experimental data is observed by the PHOBOS Collaboration.}
 \label{FIG4}
 \end{figure}

%
%

\end{document}